\documentclass[journal]{IEEEtran}
\usepackage{cite}
\usepackage{amsmath} 

\usepackage{subfigure}
\ifCLASSINFOpdf
\usepackage[pdftex]{graphicx}
  \graphicspath{{../pdf/}{../jpeg/}}
  \DeclareGraphicsExtensions{.pdf,.jpeg,.png}
\else
  \usepackage[dvips]{graphicx}
  \graphicspath{{../eps/}}
  \DeclareGraphicsExtensions{.eps}
\fi  
\usepackage{amsmath}
\usepackage{cases}
\usepackage{stfloats}
\usepackage{amsfonts}
\usepackage{subeqnarray}
\usepackage{longtable}
\usepackage{supertabular}
\usepackage{setspace}
\usepackage{multirow}
\usepackage{booktabs}
\usepackage[ruled,linesnumbered]{algorithm2e}
\makeatletter
\newcommand{\nosemic}{\renewcommand{\@endalgocfline}{\relax}}% Drop semi-colon ;
\newcommand{\dosemic}{\renewcommand{\@endalgocfline}{\algocf@endline}}% Reinstate semi-colon ;
\let\oldnl\nl% Store \nl in \oldnl
\newcommand{\nonl}{\renewcommand{\nl}{\let\nl\oldnl}}% Remove line number for one line
\makeatother
\usepackage[export]{adjustbox} 
\usepackage{booktabs}
\usepackage{setspace}
\usepackage{xcolor}
\usepackage{mathrsfs}
\usepackage{amsmath}
\usepackage{array}
\usepackage{amssymb}
\usepackage{amsthm}
\usepackage{microtype}
\usepackage{url}
\usepackage{amsfonts,amssymb}
\usepackage{dsfont}
\usepackage{mathtools}
\usepackage{xcolor,colortbl}
\usepackage{colortbl}
\usepackage{graphicx}
\usepackage{enumitem}
\usepackage[utf8]{inputenc}
\usepackage{booktabs}

\definecolor{Gray}{gray}{0.85}
\definecolor{Whitecolor}{rgb}{1,1,1}
\definecolor{cincinnati-red}{RGB}{190,0,0}
\definecolor{purpple}{RGB}{251,142,50}

\newcolumntype{a}{>{\columncolor{Gray}}c}
\newcolumntype{b}{>{\columncolor{white}}c}
\allowdisplaybreaks
\begin{document}
\setstretch{1}
\title{\textls[-25]{Data-driven Coordinated AC/DC Control Strategy for Frequency Safety}}
\author{Qianni~Cao,~\IEEEmembership{Student~Member,~IEEE},
Chen~Shen,~\IEEEmembership{Senior~Member,~IEEE}

%\thanks{M. Jia and C. Shen are with the State Key Laboratory of Power Systems, Tsinghua University, 100084 Beijing, China. Y. Wang and G. Hug are with the Power Systems Laboratory, ETH Zurich, 8092 Zurich, Switzerland.}
}
        
%\thanks{This work was supported in part by the Joint Funds of the National Natural Science Foundation of China under Grant U1766206 (Correspondence to Chen Shen).}
%\thanks{M. Jia, C. Shen and Z. Wang are affiliated with the State Key Laboratory of Power Systems, Department of Electrical Engineering, Tsinghua University, Beijing 100084, China (e-mail addresses: jms16@mails.tsinghua.edu.cn, shenchen@mail.tsinghua.edu.cn,
%    wangzhaojian@mail.tsinghua.edu.cn).}% <-this % stops a space
% \thanks{Manuscript received April 19, 2005; revised August 26, 2015.}

%\markboth{Submitted to IEEE Trans. Smart Grid}%
%{Shell \MakeLowercase{\textit{et al.}}: Bare Demo of IEEEtran.cls for IEEE Journals}
\maketitle

\begin{abstract}
With high penetrations of renewable energy and power electronics converters, less predictable operating conditions and strong uncertainties in under-frequency events pose challenges for emergency frequency control (EFC). On the other hand, the fast adjustability of converter-based sources presents opportunities to reduce economic losses from traditional load shedding for EFC. By integrating DC power emergency support, a data-driven coordinated AC/DC control strategy for frequency safety - Coordinated Emergency Frequency Control (CEFC) - has been designed. CEFC coordinates both the initiation and control amount of emergency DC power support (EDCPS) and traditional load shedding. Based on real-time power system response data, CEFC ensures system frequency safety at a minimum control cost under non-envisioned operating conditions and large power deficits. A sufficient condition where data-driven modeling errors do not affect the precision of the control strategy for power system frequency is rigorously provided. Simulation results demonstrate CEFC's adaptability, prediction accuracy, and control effectiveness.
\end{abstract}
%Although the historical data of renewable generations could be assumed as publicly known
% Note that keywords are not normally used for peerreview papers.
\IEEEpeerreviewmaketitle

% \section*{Nomenclature}
% \addcontentsline{toc}{section}{Nomenclature}

% \subsection*{Indices} 
% \begin{IEEEdescription}[\IEEEusemathlabelsep\IEEEsetlabelwidth{$aaaaaaaa$}]

% \end{IEEEdescription}

\section{Introduction}
\subsection{Motivation}
\IEEEPARstart{W}{ith} the integration of renewable energies and development of HVDC transmission technologies, traditional AC grids have transformed into hybrid AC-DC grids with increased renewable energy generation. In AC-DC systems, DC block faults or AC-DC cascading faults are prone to occur, which could cause a considerable active power imbalance \cite{9328133}. Moreover, the declining system inertia leads to too low a frequency nadir which cause serious consequences \cite{4495556,Bevrani2009}. Therefore, modern power systems require an emergency frequency control (EFC) strategy to deal with the considerable power imbalance.

% Traditional EFC strategies typically fall into one of two categories: offline decision-making real-time matching strategies, or online decision-making real-time matching strategies. Both approaches rely on the anticipation of faults, with the former also considering anticipated operating conditions. For the sake of clarity, we will refer to both as pre-scheduled EFC hereafter. However, in modern power systems, the fluctuating of renewable energy generation results in less predictable operating conditions and strong uncertainties in under frequency events. As a consequence, traditional pre-scheduled EFC strategies become less effective\cite{10375966}.

Traditional EFC strategies are pre-scheduled, relying on offline simulations with anticipated operating conditions and faults. However, in modern power systems, less predictable operating conditions and strong uncertainties in under frequency events make these strategies less effective. Furthermore, the traditional approaches of EFC are generator tripping or load shedding operations \cite{1137600,5546962}, but these operations cause severe economic losses. With more inverter-connected resources and their fast adjustability to provide frequency support, a coordinated EFC approach that combines emergency DC power support (EDCPS) with load shedding can be developed to enhance the stability of system frequency. The coordination of traditional load shedding and EDCPS offers a more economically viable solution to maintaining frequency stability.

In this paper, a real-time decided and non-preplanned EFC scheme, combining and coordinating load shedding and EDCPS, is designed to promise frequency safety. With EDCPS, the scheme reduces or replaces load shedding in under-frequency events. Firstly, we realized system identification under multiple control resources with offline data. Then, the frequency nadir and steady-state value is predicted when EDCPS is implemented. If the predicted system frequency falls below a safety range, a one-shot load shedding is initiated. We then investigate how inaccuracies in system identification affect the control strategy. This approach does not depend on anticipated operating conditions and faults, thus potentially preventing the possibility of under or over-shedding that may occur with traditional pre-scheduled schemes.

\subsection{Literature Review}
At present, the majority of relevant research focuses on EFC with a single control resource, such as emergency load shedding, EDCPS or Energy Storage Systems (ESS). Conventional EFC are stepwise load shedding processes based on prescheduled tables that are determined based on a rule of experts and past experience\cite{vittal2019power}. This approach requires extensive offline simulation and thus suffers from low efficiency.

To avoid extensive offline simulation, methods utilizing classical models of center of inertia (COI) frequency dynamics have been introduced. Ref. \cite{7581062} determines the minimum load to shed post-disturbance based on the swing equation, considering the ramp-up limits and capacities of committed units. Ref. \cite{10151932} presents an analytical expression for the optimal control strategy of ESS to ensure adequate frequency support, based on the swing equation and first-order Primary Frequency Response (PFR) dynamics. These classical representations are used to set constraints on system frequency dynamics, forming an optimal frequency control framework with clear physical interpretations. However, parameters in these classical power system representations, like the system inertia and damping coefficients, are not directly measurable and can be difficult to estimate accurately. Additionally, these parameters may change under events of emergency faults.

To address the limitations of control strategies based on traditional power system models, recent research has begun to integrate data-driven approaches with classical model-based approached. Ref. \cite{9767658} leverages system measurements to refine the parameters of these traditional models. Yet, the effectiveness of this approach is constrained by the intrinsic accuracy of the classical models. Ref. \cite{8723612} introduces neural networks to enhance or correct the predictions of classical models. However, representing the system's input-output relationship with neural networks as constraints in an optimization problem poses challenges for straightforward application with commercial optimizers, hindering online application.

Bypassing physical models, Ref. \cite{7337458} predicts system frequency a few seconds ahead by fitting data to a second-order polynomial. This approach is seen as more adaptable to variations of operating conditions and power deficits. However, its focus on local trajectory predictions requires constant updates, optimizing the control strategy at every time step. Therefore, it suits control resources that allow frequent adjustments but is impractical for those that cannot be activated regularly.

Unlike local trajectory predictions, Koopman Operator (KO) Theory aims to globally linearize nonlinear dynamics by elevating the state space to a higher-dimensional observable space \cite{koopman1931hamiltonian}. Koopman-operator-based representations offers greater flexibility than traditional models of power systems \cite{KORDA2018297}. Meanwhile, linear time-invariant equations allow the use of efficient linear control tools. However, the application of KO theory in engineering faces several challenges \cite{doi_10_1137_21M1401243}. Firstly, constructing observation functions demands carefully design of their basis functions, and incorrect basis function choices can hinder attaining even basic models \cite{Kaiser_2021}. Additionally, approximation errors of Koopman operators may arise from truncation, sampling bias and training errors\cite{cao2023data}. Thus, understanding the effects of these inaccuracies on control strategies is crucial for leveraging Koopman theory effectively. 
In summary, a major challenge lies in modeling frequency dynamics and understanding how modelling inaccuracies can affect control strategies.
%understanding how approximation errors across diverse operating conditions, non-envisioned disturbances, and various control strategies over a relatively long term until frequency stabilizes.

Currently, the research for coordinating multiple control resources in emergency frequency control are in early development. Ref. \cite{xu2017design} suggests a framework that integrates DC power emergency support, pumped-storage plants, and load shedding, providing guidelines for resource coordination. Ref. \cite{10371235} introduces a P-F droop strategy for distributed generators and a load shedding plan to tackle power deficits and maintain stability. Ref. \cite{santurino2022optimal} formulates an offline nonlinear optimization problem focused on representative contingencies, aiming to fine-tune frequency thresholds, shedding levels, and the droop and virtual inertia coefficients for ESS. However, these methods are still constrained by classical representations or anticipated contingencies.

Coordinating multiple control resources requires considering the distinct properties of each resource, such as control cost and flexibility. This paper primarily investigates two types of control resources: load shedding (a traditional control method) and DC power regulation (an inverter-based resource). Load shedding is highly undesirable for electricity consumers, yet traditional methods often lead to excessive shedding due to step-by-step protection actions and inherent time delays \cite{7581062}. In hybrid AC-DC systems, DC block faults can cause a greater active power deficit than in conventional systems, making over shedding even more unwanted \cite{9328133}. Thus, developing a one-shot load shedding scheme becomes crucial, acting as an open-loop control mechanism \cite{7581062}. On the other hand, adjusting the DC power reference offers a fast, adaptable resource, allowing continuous control adjustments based on system measurements until frequency recovery \cite{10049715}. This adjustment represents a closed-loop control strategy. 

Coordinating diverse control resources presents itself as another key challenge. On one hand, the DC power reference is adjusted adaptively throughout the entire control process, whereas load shedding is a one-time action. Therefore, it becomes challenging to accurately forecast the strategy of the DC power when the load shedding strategy is decided. On the other hand, unlike DC power's closed-loop control that enables continuous strategy adjustments over a receding horizon, load shedding, as a one-time action, demands greater precision in control strategy. The impact of system dynamic modeling errors on the switching of load feeders constitutes a research gap.

\subsection{Contribution}
According to the literature review, to address the emergency frequency instability problems with multiple control resources and close the aforementioned gaps, the contributions of this paper are as follows:
\begin{enumerate}
\item A modelling method is developed with global accuracy to predict the system frequency. The method is adaptive to non-envisioned operating conditions, unanticipated power deficit and different types of control resources. 

\item A fully data-driven emergency control method is designed, which coordinates DC power modulation and load shedding.

\item The robustness of the proposed method against system identification errors is analyzed, deriving sufficient conditions under which system identification errors do not affect the results of the load shedding.
\end{enumerate}

The rest of this paper is organized as follows. Section II develops a system frequency modelling methods considering different control resources. Based on it, we proposes the coordination strategy for under frequency events. Section III investigates how approximation inaccuracies in the Koopman linear representations influence the results of one-shot load shedding. In Section IV, an CIGRE-FS system case is presented and the effectiveness of the proposed control strategy is verified. Section V provides the conclusion.

\section{Emergency Frequency Controller Design}

In this section, we develop a data-driven system frequency modelling methods with global accuracy. It provides the constraints on system dynamics in an optimal frequency control framework. Then, a EFC optimal control strategy which coordinates load shedding and DC power reference is developed.

\subsection{System frequency modelling}

This section proposes a method as a data-driven approximation of KO. The key idea is to globally linearize the nonlinear dynamics by lifting the state space to a hyperdimensional observable space.

Based on Koopman theory, we assume that the frequency dynamics of the system is governed by the linear dynamic system equation represented in Eq.\eqref{eq_linear_dynamic_system_equation}.
\begin{align}
    \boldsymbol{g}_{t+1}=\boldsymbol{A}{\boldsymbol{g}_{t}}+\begin{bmatrix}
\boldsymbol{B}_l & \boldsymbol{B}_d
\end{bmatrix}
\begin{bmatrix}
\boldsymbol{u}_{l,t} \\
\boldsymbol{u}_{d,t}
\end{bmatrix}\label{eq_linear_dynamic_system_equation}
\end{align}
where 
\begin{align}
    \boldsymbol{g}_{t}=\left[ \begin{matrix}
   \omega_{t}  \\
   \boldsymbol{\varphi} ({\omega }_{t-\tau :t},\boldsymbol{y}_{t-\tau :t})  \\
\end{matrix} \right] \label{time_delay}
\end{align}
%加一段g为什么这样构建的
where $\omega_t$ is the deviation of center of inertia (COI) frequency from its nominal value (in per unit) at time $t$, ${{\omega }_{t-\tau :t}}=[{{\omega }_{t-\tau }},{{\omega }_{t-\tau +\Delta t}},...,{{\omega }_{t}}],{\boldsymbol{y}_{t-\tau :t}}=[{\boldsymbol{y}_{t-\tau }},{\boldsymbol{y}_{t-\tau +\Delta t}},...,{\boldsymbol{y}_{t}}]$ represent the time series of system frequency (state variable) and voltages (algebraic variables), respectively. $\boldsymbol{\varphi}$ denotes a neural network with a prescribed structure. $\boldsymbol{g}$ is a set of finite Koopman observables which forms a subspace of the infinite dimensional Koopman observables. $\boldsymbol{u}$ represents the control inputs, where $\boldsymbol{u}_{d,t}$ denote the deviations of the DC power references from their pre-fault set values for LCC-HVDCs. $\boldsymbol{u}_{l,t}$ represents the dynamic load shedding ratios and static load shedding ratios at different load nodes. Following the loss function and the algorithm provided in Ref. \cite{cao2023data}, it is feasible to approximate the parameters of $\boldsymbol{g}$, along with the matrices $\boldsymbol{A}$ and $\boldsymbol{B}=[\boldsymbol{B}_l^{\top}\ \boldsymbol{B}_d^{\top}]^{\top}$. $\boldsymbol{B}_l$ and $\boldsymbol{B}_d$ represents the control matrix for load shedding and EDCPS, respectively. 

After the parameters of $\boldsymbol{g}$, $\boldsymbol{A}$, and $\boldsymbol{B}$ have been approximated from data, one can utilize Eq. \eqref{eq_linear_dynamic_system_equation} to predict the future trajectory of the system frequency variation, provided the control sequence $\boldsymbol{u}_t$, a time series of ${\omega }_{1-\tau :1}^{\alpha,\beta}$ and $\boldsymbol{y}_{1-\tau :1}^{\alpha,\beta}$. Herein, we refer to the dynamic system described by Eq. \eqref{eq_linear_dynamic_system_equation} as a Koopman linear system.

Using the linear prediction system from Eq. \eqref{eq_linear_dynamic_system_equation}, it can be predicted if direct current regulation resources can prevent the system frequency from dropping below a set threshold.

\subsection{Koopman-Operator-Based EFC Optimal Coordination Methods}
\label{Emergency_Frequency_Controller_Design}

When an under-frequency event occurs, it is necessary to coordinate two types of control resources to ensure that the system frequency remains within a safe range. According to the grid code, it is generally required that the post-event frequency nadir and the steady-state frequency  remain within a safety range.
As load shedding is undesirable for electricity consumers, the coordination principle between DC power support and load shedding aims to maximize the utilization of DC power resources to minimize the need for load shedding.

Therefore, this subsection first predicts whether the system frequency remains within the safe range under maximal utilization of DC power resources. If within the safe range, we optimizes the active power reference for each LCC-HVDC in a moving horizon fashion to provide emergency frequency support. If outside the safe range, an optimal one shot load shedding strategy is designed.

\subsubsection{Activation of EFC}

While EFC works only in the case of emergency frequency problems, a dead zone setting for both EDCPS and load shedding is necessary. There are two common methods to set the dead zone, i.e., the frequency deviation limitation and the frequency change rate limitation, and in this paper, we use the former. When the system frequency changes due to some faults, the frequency limitation of the dead zone is utilized to determine whether there is an emergency and whether to enable the EFC.

Usually, the grid operator would like to set the trigger value for DC power support higher than the threshold value for under-frequency load shedding.
%considering emergency frequency drops, the grid would like to give priority to DC power support rather than load shedding to stabilize the frequency. 
Although setting the above threshold is a common practice, it is a sub-optimal measure when accurately predicting system frequency dynamics proves difficult. When the power deficit is large and DC regulation resources are insufficient, delayed load shedding requires shedding more load, leading to greater economic losses. Since Eq.\eqref{eq_linear_dynamic_system_equation} predicts system frequency under non-envisioned operating conditions and faults, we proposes triggering the EDCPS once frequency drops to a preset threshold, and at the same time calculating and executing the optimal load shedding amount if needed. As Eq.\eqref{eq_linear_dynamic_system_equation} is linear, applying it for frequency prediction and as a constraint for optimal control is time-efficient. Therefore, this data-driven load shedding generally occurs before the system frequency hits the trigger value for load shedding specified in grid code "DL/T 428-2010 \cite{DL428-2010}" after a disturbance. Earlier load shedding reduces the amount required to maintain frequency within the safe range. Thus, the proposed method sheds less load than traditional UFLS.

In summary, EDCPS is initiated when the system frequency drops to a preset value, simultaneously triggering the determination of whether load shedding is required. When load shedding is necessary, the shedding occurs slightly after EDCPS initiation, with the delay accounting for the time to determine the need for shedding and calculate the optimal shedding amount. First, the load shedding initiation method will be introduced, followed by the optimal load shedding calculation method, and finally, the design of the DC control strategy.

\subsubsection{Initiation for load shedding} 
\label{Initiation_for_load_shedding}

Based on the system frequency dynamics, we first predicts the system frequency under maximal utilization of DC power resources. We assume that $\boldsymbol{u}_{d,t} (t=1,2,…,T)$ takes the value of the upper limit of DC active power for the sending end (and the lower limit for the receiving end), while $\boldsymbol{u}_{l,t}  (t=1,2,…,T)$ is set as a zero vector. These values are substituted into Eq. \eqref{eq_linear_dynamic_system_equation}. If the system's lowest frequency is predicted to be higher than the predetermined threshold, it indicates that by regulating the DC power can ensure system safety without load shedding. 

If the frequency nadir falls below the threshold, emergency DC support may not guarantee system safety, necessitating load shedding measures for frequency stability. If load shedding is required, proceed to calculating the optimal amount for load shedding; if not, proceed to calculating the optimal amount for DC power regulation.

\subsubsection{Load shedding strategy}

If it can be anticipated that, under emergency DC support, the frequency nadir falls below a preset threshold, a data-driven one-shot load shedding can be triggered. 

In engineering practice, EFC typically involves a main station that collects measurements from substations. The main station then calculates and allocates the shedding amount to each substation. Subsequently, each execution station distributes the shedding amount to individual feeders. Following the engineering practice, we first calculate the load shedding amount for each substation, and then rounded them to the nearest discrete value.

The shedding amount is determined by solving the following optimal control problem.
\begin{align}
&\ \ \ \ \ \ \ \ \ \min \ \sum_{t=1}^{T} \boldsymbol{u}_t^{\top}\boldsymbol{Q}_1 \boldsymbol{u}_t \notag\\
&\text{ s.t. }\ \ \boldsymbol{g}_{t+1} = \boldsymbol{A}\boldsymbol{g}_t + \begin{bmatrix} \boldsymbol{B}_l & \boldsymbol{B}_d \end{bmatrix} \begin{bmatrix} \boldsymbol{u}_{l,t} \notag\\ \boldsymbol{u}^{MAX}_{d,t} \end{bmatrix}, \forall t = 1,2, \ldots, T \notag\\
&\ \ \ \ \ \ \ \omega_t \geq \omega_{\min}, \forall t = 1,2, \ldots, T \label{optimal_load_shedding}\\
&\ \ \ \ \ \ \ \boldsymbol{u}_{l,2} = \boldsymbol{u}_{l,3} = \cdots = \boldsymbol{u}_{l,T} \notag\\
&\ \ \ \ \ \ \ \boldsymbol{u}_{l,t}\le\boldsymbol{u}^{MAX}_{l,t} \notag
\end{align}
where $\boldsymbol{Q}_1$ represents a positive definite matrix used to represent the cost of regulating DC power, the parameters in $\boldsymbol{g}$, $\boldsymbol{A}$ and $\boldsymbol{B}$ are obtained offline, $\boldsymbol{u}^{MAX}_{d,t}(t = 2,3,...,T)$ is the DC power adjustable limit (upper limit for the receiving end and lower limit for the sending end), $\boldsymbol{u}_{l,1}$ is a zero vector, $\boldsymbol{u}_{l,t}(t = 2,3,...,T)$ and $\boldsymbol{g}_{t}(t = 2,3,...,T)$ are the decision variables, $\boldsymbol{u}^{MAX}_{l,t}$ denotes the maximum allowable load shedding amount at different nodes.

In practice, continuous adjustment of load shedding value is difficult to achieve, and it is often necessary to choose whether or not to shed a load on a particular feeder line, resulting in a series of discrete values for the actual load shedding. Let the optimal load shedding value obtained by solving Eq.\eqref{optimal_load_shedding} be denoted as ${\boldsymbol{\bar{u}}_{l*}}=[{{\bar{u}}_{l,1*}},...{{\bar{u}}_{l,i*}}...,{{\bar{u}}_{l,I*}}]$ the actual load shedding amount would then be given as

\begin{align}
    {{Q}_{d}}\left( {{{\bar{u}}}_{i*}} \right)=\left\{ \begin{matrix}
   nd & nd\le {{{\bar{u}}}_{i*}}<n\left( d+0.5 \right)  \\
   n\left( d+1 \right) & n\left( d+0.5 \right)\le {{{\bar{u}}}_{i*}}<n\left( d+1 \right)  \\
\end{matrix} \right. \label{actual_load_shedding_amount}
\end{align}
where $d$ represents the quantization interval, which physically refers to the load shedding amount on a single feeder line, $n$ represents a positive integer, and $Q_d({\bar{u}}_{i\ast})$ denotes the actual load shedding amount at each load node $i$ when the discrete interval is $d$.

%The reason for initially calculating the shedding amount as a continuous quantity and then rounding it to the nearest value lies in the fact that EFC typically involves a main station collecting measurements from substations, and then allocating the shedding amount to each substation. Subsequently, each execution station distributes the shedding amount to individual feeders. The process of Eq. \eqref{optimal_load_shedding} and Eq. \eqref{actual_load_shedding_amount} which first calculate a continuous shedding amount and then rounding it aligns with the actual decision-making process in engineering applications.

Eq. \eqref{optimal_load_shedding} and Eq. \eqref{actual_load_shedding_amount} also circumvents the NP-hard problem arising from solving 0-1 optimization problems, hindering the computational efficiency of online optimization.

It is noteworthy that the optimal load shedding problem Eq. \eqref{optimal_load_shedding} is linear, and thus can be solved in polynomial time. If the computation time for solving the optimal load shedding amount is typically within $\Delta t$, the prediction time step in Eq.\eqref{optimal_load_shedding} can be set greater than or equal to $\Delta t$. The optimal load shedding control action is applied to the system at the second time step.

% If the computation time for solving the optimal load shedding amount is typically within $\Delta t$, the prediction time step in Eq.\eqref{optimal_load_shedding} is $\Delta t$,  Therefore, the optimal load shedding control action is generally applied to the system at $t=1$.

\subsubsection{Optimal DC power reference regulation}

In this subsection, we develop a data-driven closed-loop strategy, which optimizes the active power reference for each LCC-HVDC in a moving horizon fashion to provide emergency frequency support.

Instead of operating the active power at maximum capacity, a closed-loop control strategy for LCC-HVDC systems is necessary under two circumstances: i) When load shedding is unnecessary, we need to minimize the control cost for DC power regulation while maintaining frequency stability; ii) When load shedding is necessary, a closed-loop control strategy effectively reduces DC regulation amount as frequency recovers, improving cost efficiency.
%When the frequency nadir violates the minimum threshold but the steady-state value remains within the prescribed range, a closed-loop control strategy helps to reduce the DC power reference regulation and lower the control cost.
%The closed-loop strategy在两种情况下是需要的，第一种情况是当系统有功缺额不大，无需切负荷时，需要最小化DC power support的成本，另一种情况是存在load shedding的情况下，系统频率回升，DC power support无需一直维持在上限值（或下限值）。

In general, the active power regulation can be achieved by solving a discrete Linear Quadratic Regulator (LQR) problem as given in Eq. \eqref{Koopman_based_LQR_for_HVDC}. The robustness of LQR against Koopman-based modeling errors in system characteristics has been proved in Ref. \cite{cao2023data}.

\begin{align}
&\min \sum_{t=1}^{T} \boldsymbol{g}_t^\top \boldsymbol{Q}_2\boldsymbol{g}_t+ \boldsymbol{u}_t^\top \boldsymbol{R}_2\boldsymbol{u}_t \notag\\
&\text{s.t.}\quad \boldsymbol{g}_{t+1} = \boldsymbol{A}\boldsymbol{g}_t + \begin{bmatrix} \boldsymbol{B}_l & \boldsymbol{B}_d \end{bmatrix} \begin{bmatrix} \boldsymbol{u}_{l,t} \\ \boldsymbol{u}_{d,t} \end{bmatrix}, \quad \forall t = 1,2,\ldots,T \label{Koopman_based_LQR_for_HVDC}
\end{align}
where $\boldsymbol{Q}_2\ge0,\boldsymbol{R}_2>0$ are diagonal weight matrix minimizing deviations from the desired frequency and the control sost, respectively, $\boldsymbol{u}_{l,t}$ are the zero vector when load shedding is not required, $\boldsymbol{u}_{d,t}(t = 2,3,...,T)$ and $\boldsymbol{g}_{t}(t = 2,3,...,T)$ are the decision variables.
The optimum of Eq.\eqref{Koopman_based_LQR_for_HVDC} can be solved by a discrete time algebraic Riccati equation (DT ARE) given as 
\begin{align}
\boldsymbol{u}^{*}_{d,t} = -(\boldsymbol{R}_2 + \boldsymbol{B}_d^{\top} \boldsymbol{P} \boldsymbol{B}_d)^{-1} \boldsymbol{B}_d^{\top} \boldsymbol{P} \boldsymbol{A} \boldsymbol{g}_k
\end{align}
where $\boldsymbol{P}$ satisfies
\begin{align}
\boldsymbol{A}^{\top} \boldsymbol{P} \boldsymbol{A} &- \boldsymbol{P} + \boldsymbol{Q}_2 &  \notag \\- &\boldsymbol{A}^{\top} \boldsymbol{P} \boldsymbol{B}_d (\boldsymbol{R}_2 + \boldsymbol{B}_d^{\top} \boldsymbol{P} \boldsymbol{B}_d)^{-1}\boldsymbol{B}_d^{\top} \boldsymbol{P} \boldsymbol{A} = \boldsymbol{0}
\end{align}

If load shedding is initiated, $\boldsymbol{u}_{l,t}$ takes the value calculated by Eq.(3)-(4). Then Eq.\eqref{Koopman_based_LQR_for_HVDC} can be solved efficiently by off-the-shelf commercial solvers. If $\boldsymbol{u}^{*}_{d,t}$ exceeds the adjustable limit of the DC power, it is constrained to the limit. 

To sum up, in the above Step 1)-4), by incorporating the upper or lower limits of DC power support in the system's frequency dynamics, the load shedding strategy is designed bypassing the need for predicting the exact future behavior of DC power. By using an optimization approach based on a linear frequency dynamic model and fixed DC power regulation quantities, it allows for quick calculation of load shedding amounts, enhancing calculation speed.

However, assuming the DC power regulation is always at its upper/lower limit, i.e. a step-change function can cause issues, such as mismatches between expected and actual system dynamics, leading to inaccuracies in one-shot load shedding amounts calculation.
%It aims to maximize DC regulation resource use, aligning with control coordination principles. 

The reasons why the DC regulation amount may not conform to a step-change function can arise from two reasons:
\begin{enumerate}[label=\roman*)]
    \item After emergency control commands are issued, the real-time support power of the DC line may not quickly adjust to these commands.
    \item With the use of a LQR strategy for frequency control, the amount of DC active power regulation might decrease as frequency stabilizes, leading to a deviation from a step-change function.
\end{enumerate}

The load shedding scheme in this paper is designed as a one-shot process, which demands high precision in its control strategy. Therefore, the following section will analyze how errors in Koopman-based system dynamic modeling affect the switching of load feeders.

\section{Robustness Analysis and Error Estimation}

% Unlike DC power’s closed-loop control that enables continuous strategy adjustments over a receding horizon, load shedding, as a one-time action, demands greater precision in control strategy.
In this subsection, we analyze the impact of system dynamic modeling errors on the switching of load feeders. This subsection provides a sufficient condition to ensure that system dynamic modeling errors do not cause the load feeder switching obtained in Section \ref{Emergency_Frequency_Controller_Design} to deviate from the optimal switching. It mathematically proves the existence of a condition on modeling errors that enables optimal one-shot load shedding, ensuring the validity of the proposed load shedding method.
%具体而言，给出了system dynamic modeling errors不导致上一小节求得的 switching of load feeders 不同于最佳switching of load feeders的充分条件。该条件mathematically证明了，尽管system dynamic modeling errors不可避免，但是存在一个system dynamic modeling errors的条件，当该条件满足时，optimal one-shot load shedding can be achieved，从而保证了第二节提出的切负荷方法的使用条件。

As analyzed in subsection \ref{Emergency_Frequency_Controller_Design}, dynamic modeling errors result from two sources: prediction errors in DC power support, and the inevitable training errors and sampling bias associated with Koopman representations. This subsection proposes a generalized methodology to simultaneously account for the impact of dynamic modeling errors originating from different sources on load shedding outcomes.

If there are $N$ feeder lines, the number of potential ways to shed these lines is $\mathcal{N}=i^{N}$, yielding $\mathcal{N}$ time-invariant systems.

Consider the linear system which can switch between the following $\mathcal{N}$ time-invariant modes $\{S_i:\boldsymbol{A}\boldsymbol{g}(\textbf{\textit{x}})+\boldsymbol{B}_i|i=1,2,\cdots,\mathcal{N}\}$. The switched system can be described by the form
\begin{align}
    \boldsymbol{g}(\textbf{\textit{x}}_{t+1})&=f(\textbf{\textit{x}}_{t},\boldsymbol{v})\notag\\
    &=(\boldsymbol{A}\boldsymbol{g}(\textbf{\textit{x}}_{t})+\boldsymbol{B}_1)v_1+(\boldsymbol{A}\boldsymbol{g}(\textbf{\textit{x}}_{t})+\boldsymbol{B}_2)(1-v_1)v_2 \notag\\
    &+\cdots+(\boldsymbol{A}\boldsymbol{g}(\textbf{\textit{x}}_{t})+\boldsymbol{B}_{\mathcal{N}-1})\prod_{i=1}^{\mathcal{N}-2}(1-v_{i})v_{\mathcal{N}-1}\notag\\
    &+(\boldsymbol{A}\boldsymbol{g}(\textbf{\textit{x}}_{t})+\boldsymbol{B}_\mathcal{N})\prod_{i=1}^{\mathcal{N}-1}(1-v_i) \label{eq_learnt_nonlinear_system}
\end{align}
where
\begin{align}
    \boldsymbol{B}_{i}=\boldsymbol{B}\boldsymbol{u}^{i}
\end{align}
where $\boldsymbol{u}^{i}$ represents $i$-th mode to shed the feeders, $v_{i}$ represent $i$-th time-invariant mode. In the $i$-th load shedding mode, $v_{i}=1, v_{j}=0(j\neq i)$.

\vspace{0.2cm}
\noindent \textbf{Proposition 1.} The linear system which can switch between the $\mathcal{N}$
time-invariant modes in $\{S_i:\boldsymbol{A}\boldsymbol{g}(\textbf{\textit{x}})+\boldsymbol{B}_i|i=1,2,\cdots,\mathcal{N}\}$ can be described by Eq. \eqref{eq_learnt_nonlinear_system} with admissible inputs $\boldsymbol{v}_i\in\{0, 1\}$, $i=1,2,\cdots,\mathcal{N}-1$. The prerequisite condition for ensuring that the representation error does not influence the load shedding results is characterized by the condition:
\begin{align}
    k^{*}=i^{*}
\end{align}
% \begin{align}
%     \lambda(t)^\top(\boldsymbol{A}\boldsymbol{g}(\textbf{\textit{x}})+\boldsymbol{B}_{i^{*}})+\boldsymbol{P}_{i^{*}}<\lambda(t)^\top(\boldsymbol{A}\boldsymbol{g}(\textbf{\textit{x}})+\boldsymbol{B}_{k^{*}})+\boldsymbol{P}_{k^{*}}
% \end{align}
where
% \begin{align}
%     \gamma^{\top}(t)(\boldsymbol{\varepsilon_{A}}\boldsymbol{g}&(\textbf{\textit{x}})+\varepsilon_{\boldsymbol{B}k*})\notag\\
%     &<\gamma^{\top}(t)(\boldsymbol{A}\boldsymbol{g}(\textbf{\textit{x}})+\boldsymbol{B}_{i*})-\gamma^{\top}(t)(\boldsymbol{A}\boldsymbol{g}(\textbf{\textit{x}})+\boldsymbol{B}_{k*})
% \end{align}
\begin{align}
    k^{*}=\mathop{\text{arg min}}\limits_{k\in\{1,2,\cdots,\mathcal{N}\}} \lambda(t)^\top(\boldsymbol{A}\boldsymbol{g}(\textbf{\textit{x}})+\boldsymbol{B}_{k})+\boldsymbol{P}_k, \forall t\in [1:T] \label{argmin_eq1}
\end{align}
\begin{align}
        i^{*}=\mathop{\text{arg min}}\limits_{i\in\{1,2,\cdots,\mathcal{N}\}} \lambda(t)^\top(\bar{\boldsymbol{A}}\bar{\boldsymbol{g}}(\textbf{\textit{x}})+\bar{\boldsymbol{B}}_{i})+\boldsymbol{P}_i, \forall t\in [1:T] \label{argmin_eq2}
\end{align}
where $\boldsymbol{P}_*$ represents the control cost deduced from the control objective in Eq.\eqref{optimal_load_shedding} when $v_*=1$, $\bar{\boldsymbol{A}}$, $\bar{\boldsymbol{B}}$ and $\bar{\boldsymbol{g}}$ are the accurate Koopman linear representations, which accurately captures the system dynamics, $\lambda(t)\in \mathbb{R}^{dim(\boldsymbol{g})}$ are the solution of 
\begin{align}
    \lambda(t+1)&=-\boldsymbol{A}^{\top}\lambda(t) [v_1+(1-v_1)v_2 \notag\\
    &+\cdots+\prod_{i=1}^{\mathcal{N}-2}(1-v_{i})v_{\mathcal{N}-1}+\prod_{i=1}^{\mathcal{N}-1}(1-v_i) ].
\end{align}
with the boundary condition $\lambda(T)=0$.

\vspace{0.2cm}
\noindent \textbf{Proof.} Consider the dynamical system in Eq. \eqref{eq_learnt_nonlinear_system}, we relaxed the constraint on the system mode and allowed them to vary continuously between 0 and 1, i.e., $v_{i}\in [0,1]$.
By augmenting the cost functional $J$ in Eq. \eqref{optimal_load_shedding} with the constraint in Eq. \eqref{eq_learnt_nonlinear_system}, we have
% \begin{align}
% J &= \sum_{t=1}^{T} \bigg[f_0(k, x_k, u_k) + \gamma^{\top}\big(x_{k+1} - f(k, x_k, u_k)\big)\bigg]
% \end{align}
\begin{align}
J &= \sum_{t=1}^{T-1} \bigg[\boldsymbol{u}_t^{\top}\boldsymbol{Q}_1 \boldsymbol{u}_t + \lambda^{\top}\big(\boldsymbol{g}(\boldsymbol{x}_{t+1}) - f(\textbf{\textit{x}}_{t},\boldsymbol{V}_{t})\big)\bigg]
\end{align}
We define the the Hamiltonian as follows
\begin{align}
\boldsymbol{H}(t, \textbf{\textit{x}}_{t},\boldsymbol{V}_{t},\lambda) &= \boldsymbol{Q}_1 \boldsymbol{u}_t + \lambda^{\top} f(\textbf{\textit{x}}_{t},\boldsymbol{V}_{t})
\end{align}

To determine the optimal control inputs, $\boldsymbol{u}^*_{l,t}$, we apply the Pontryagin’s Minimum Principle to get
\begin{align}
    \boldsymbol{H}(t, \textbf{\textit{x}}^{*}_{t},\boldsymbol{V}^{*}_{t},\lambda)  \le \boldsymbol{H}(t, \textbf{\textit{x}}_{t},\boldsymbol{V}_{t},\lambda) \ \ \  \forall t = 1,2,...,T \label{Hamiltonian}
\end{align}

By inspecting Eqs. \eqref{eq_learnt_nonlinear_system} and \eqref{Hamiltonian}, we deduce that we effectively need to minimize
\begin{align}
    \boldsymbol{H}(t, &\textbf{\textit{x}}_{t},\boldsymbol{V}_{t}, \lambda)  =\lambda^{\top}[(\boldsymbol{A}\boldsymbol{g}(\textbf{\textit{x}}_{t})+\boldsymbol{B}_1+\lambda^{-\top}\boldsymbol{P}_1)v_1 \notag\\
    &+(\boldsymbol{A}\boldsymbol{g}(\textbf{\textit{x}}_{t})+\boldsymbol{B}_2+\lambda^{-\top}\boldsymbol{P}_2)(1-v_1)v_2+\cdots \notag\\
    % &+(\boldsymbol{A}\boldsymbol{g}(\textbf{\textit{x}}_{t})+\boldsymbol{B}_{\mathcal{N}-1}+\lambda^{-\top}\boldsymbol{P}_{\mathcal{N}-1})\prod_{i=1}^{\mathcal{N}-2}(1-v_{i})v_{\mathcal{N}-1}\notag\\
    &+(\boldsymbol{A}\boldsymbol{g}(\textbf{\textit{x}}_{t})+\boldsymbol{B}_\mathcal{N}+\lambda^{-\top}\boldsymbol{P}_{\mathcal{N}}))\prod_{i=1}^{\mathcal{N}-1}(1-v_i)] \label{minimizing}
\end{align}
fot $t = 1,2,...T$, with respect to $v_i, i= 1,2,...,(\mathcal{N}-1)$, Eq. \eqref{minimizing} can be expressed as
\begin{align}
    \bar{\boldsymbol{H}}(t, &\textbf{\textit{x}}_{t},\boldsymbol{V}_{t}, \lambda)=a_{1}w_{1}+a_{2}w_{2}+...+a_{N}w_{N}
\end{align}
where
\begin{align}
    a_{i}=\lambda^{\top}(\boldsymbol{A}\boldsymbol{g}(\textbf{\textit{x}}_{t})+\boldsymbol{B}_i)+\boldsymbol{P}_i
\end{align}
\begin{align}
    w_{i} = v_{i} \prod_{n=1}^{i-1} (1 - v_n) \label{w_v}
\end{align}
for $i = 1,2,...,N,$.

% \begin{align}
%     &w_{1} = v_{1}\\
%     &w_{2} = (1-v_{1})v_{2}\\
%     &\phantom{aaa} \vdots \\
%     &w_{N-1} = v_{N-1} \prod_{i=1}^{N-2} (1 - v_i) \\
%     &w_{N} = \prod_{i=1}^{N-1} (1 - v_i). \label{w_v}
% \end{align}
From the above equation it can be shown that $w_{i}\in[0,1]$, since $v_{i}\in[0,1]$, and $\sum_{i=1}^{N} w_i = 1$. Thus,
\begin{align}
    \min \bar{\boldsymbol{H}}(t, &\textbf{\textit{x}}_{t},\boldsymbol{V}_{t}, \lambda) = a_k = \min \{a_1,a_2,...,a_N\} \label{min_hamiltonian}
\end{align} 
which is achieved by choosing $w_k = 1$ and $w_i = 0, \forall i \neq k$. According to Eq. \eqref{w_v}, this can be achieved by
\begin{align}
    v_k^* = 1, \quad v_i^* = 0, \quad v_j^* \in [0, 1], \quad \forall i \leq k, \quad \forall j \geq k
\end{align}
which includes the input combination $v_k^* = 1, v_i^* = 0, \forall i \neq k $. Although we relaxed the constraint on the control inputs and allowed them to belong to the set $[0, 1]$, optimality is achieved with all inputs taking on values of zero or unity, i.e., $ v_i^* \in \{0, 1\}, i = 1, 2, \ldots, (N - 1) $. Since $\{0, 1\} \subseteq [0, 1]$, we claim $v_i^* \in \{0, 1\}, i = 1, 2, \ldots, (N - 1)$ for the original dynamical system in Eq. \eqref{eq_learnt_nonlinear_system} with admissible inputs $ v_i \in \{0, 1\} $. Furthermore, Eq. \eqref{argmin_eq1}, \eqref{argmin_eq2} can be deduced from Eq. \eqref{min_hamiltonian} and this concludes the proof. $\square$

% \def\SetClass{article}
% \documentclass{\SetClass}
% \usepackage[ruled,linesnumbered]{algorithm2e}
% \begin{document}
% \begin{algorithm}
% \caption{Simulation-optimization heuristic}\label{algorithm}
% \KwData{current period $t$, initial inventory $I_{t-1}$, initial capital $B_{t-1}$, demand samples}
% \KwResult{Optimal order quantity $Q^{\ast}_{t}$}
% $r\leftarrow t$\;
% $\boldsymbol{\Delta} B^{\ast}\leftarrow -\infty$\;
% \While{$\boldsymbol{\Delta} B\leq \boldsymbol{\Delta} B^{\ast}$ and $r\leq T$}{$Q\leftarrow\arg\max_{Q\geq 0}\boldsymbol{\Delta} B^{Q}_{t,r}(I_{t-1},B_{t-1})$\;
% $\boldsymbol{\Delta} B\leftarrow \boldsymbol{\Delta} B^{Q}_{t,r}(I_{t-1},B_{t-1})/(r-t+1)$\;
% \If{$\boldsymbol{\Delta} B\geq \boldsymbol{\Delta} B^{\ast}$}{$Q^{\ast}\leftarrow Q$\;
% $\boldsymbol{\Delta} B^{\ast}\leftarrow \boldsymbol{\Delta} B$\;}
% $r\leftarrow r+1$\;}
% \end{algorithm}

% you can choose not to have a title for an appendix
% if you want by leaving the argument blank

\section{Case Study}
In this section, we verify the effectiveness of the proposed coordinated EFC method. First, we test the prediction accuracy of the system frequency dynamics modelling method under non-envisioned operating conditions, unpredicted disturbances and different control strategies. Then, we test if the coordinated EFC method can promise the frequency safety when i) only EDCPS is initiated and ii) when EDCPS and load shedding are simultaneously initiated.
\subsection{Test System Description}
To validate the effectiveness of our proposed method, we conducted simulation experiments on the CEPRI-FS test system. The electromechanical transient model for this case study is available for access at \cite{CSEE-Benchmark2024}. The CEPRI-FS consists of 102 500-kV buses, and possesses a load level of 2600 MW, with installed capacities of 2400 MW and 5400 MW for renewable and conventional energy sources, respectively. In the load composition, 40\% consists of the dynamic load, while 60\% is comprised of constant impedance loads. The full electromagnetic transient (EMT) model of the test system is built on the CloudPSS platform \cite{8582334}.

This study generates 300 frequency trajectories as the training set and 200 frequency trajectories as the test set for the linear dynamic system equation. The principle for generating these trajectories follows Ref. \cite{cao2023data}. In simple terms, the system's frequency dynamics are excited by i) tripping generators (up to three generators simultaneously) and ii) injecting persistently exciting white noise signals into the system (through different load nodes and DC power reference adjustments). The length of each trajectory is 60s after generator tripping. Simultaneously, the inertia of the equivalent unit is varied to simulate changes in the system's inertia caused by the commissioning and decommissioning of synchronous generators.

Frequency trajectories from the testing set are used to assess the prediction accuracy of KLS. Furthermore, 200 test scenarios are created, with the same operating conditions and faults as in the 300 trajectories. In the 200 scenarios, the control effect of different load shedding scheme is compared.

%In CEPRI-FS, Ud,t represents the deviations of the DC power references from their pre-fault set values for the 3 LCC-HVDC systems. Ul,t contains the dynamic load shedding ratios and static load shedding ratios at the 12 load nodes.

\subsection{Prediction Accuracy}

\begin{figure}[t] %可选参数 h t b p，代表允许图片出现的位置，h表示此处附近，t表示顶部，b表示底部，p表示单独一页，H表示固定此处
    \centering
    % \vspace{-0.2cm}  %调整图片与上文的垂直距离
    \setlength{\abovecaptionskip}{-0.05cm}   %调整图片标题与图距离
    \setlength{\belowcaptionskip}{-2cm}   %调整图片标题与下文距离
    \includegraphics[width=9cm]{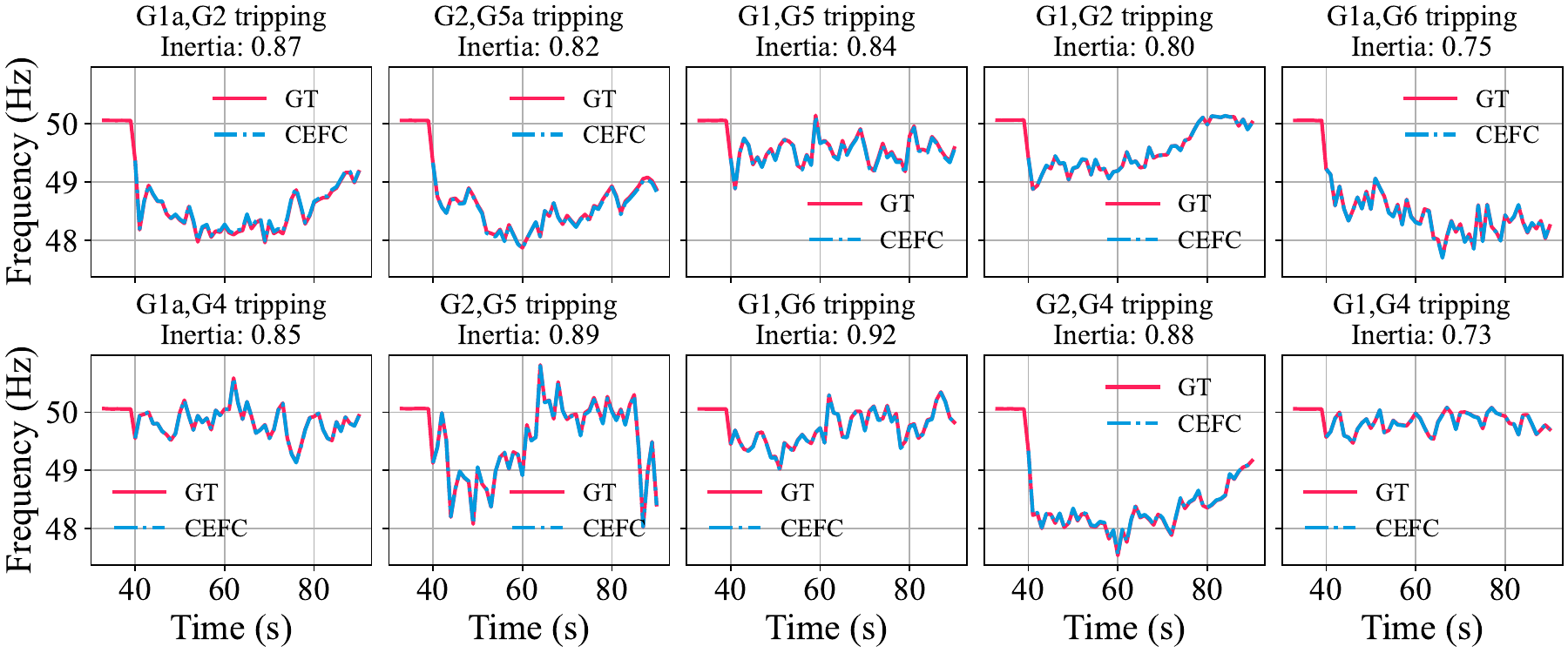}
    \caption{Comparison between true frequency trajectory(red) and predicted frequency trajectory during future 60s with proposed CEFC(blue).}\label{fig_predicted_trajectories}
\end{figure}

We test the predictive accuracy of the frequency dynamics modeling method on test datasets. Fig. \ref{fig_predicted_trajectories} shows the predictive accuracy on 10 trajectories of the test datasets, with the blue dashed line representing the frequency prediction results. The linear dynamic modelling represented in Eq. \eqref{eq_linear_dynamic_system_equation} takes the voltage at Point of Common Coupling (PCC) at find warms and the COI frequency measured 400ms after a power deficit as inputs. The red solid line represents the simulation results (ground truth). It is evident that the blue dashed line almost perfectly overlaps with the red solid line. TABLE. \ref{tab:comparison} shows the average predictive accuracy of CEFC on the trajectory of the test dataset. The table shows that the average prediction error for the frequency nadir is 0.04 Hz (which is a relative error of 0.08\% if the base value is 50 Hz), and the average prediction error for the steady-state frequency is 0.08 Hz. In Fig. \ref{fig_predicted_trajectories}  and TABLE. \ref{tab:comparison}, it can be concluded that CEFC reliably predicts frequency trajectories under non-envisioned operating conditions, unpredicted disturbances, and varying control strategies.

\begin{table}[!t]
\renewcommand{\arraystretch}{1.3} % 调整行高
\caption{The average absolute prediction errors of frequency nadir, steady-state frequency on the test dataset}
\label{tab:comparison}
\centering
\footnotesize
\begin{tabular}{@{}lccc@{}}
\toprule
Method & Nadir(Hz) & SSV(Hz) & Mean error throughout the trajectory(Hz) \\
\midrule
CEFC & 0.042 & 0.078 & 0.062 \\
CEFC-NTD & 0.288 & 0.311 & 0.286 \\
EDMD & 1.418 & 2.940 & 2.688 \\
DMD & 2.484 & 2.965 & 2.705 \\
\bottomrule
\end{tabular}
\end{table}

Furthermore, we compared the prediction errors of the proposed method with state-of-the-art algorithms (SOTAs). CEFC follows the paradigms of indirect data-driven control that first identifies a model and then conducts control design based on the identified model. Consequently, the paper compares several benchmark indirect data-driven control methods, evaluating their accuracy in system identification. We use prediction error of frequency nadir and steady-state frequency as the metric for this comparison. 

As the benchmark of the indirect data-driven control methods, the DMD and EDMD method is implemented with 100 radial basis functions (RBF) as observables. To illustrate the effectiveness of incorporating time delay embedding, CEFC without time delay embedding (CEFC-NTD, i.e., when $\tau$ = 0 in Eq. \eqref{time_delay} is also tested as a benchmark. The comparison results are given in TABLE. \ref{tab:comparison}. It can be concluded that CEFC outperforms SOTAs on prediction accuracy.

\subsection{Control Effect of Coordinated Control}
\begin{figure*}[t] %可选参数 h t b p，代表允许图片出现的位置，h表示此处附近，t表示顶部，b表示底部，p表示单独一页，H表示固定此处
    \centering
    % \vspace{-0.2cm}  %调整图片与上文的垂直距离
    \setlength{\abovecaptionskip}{-0.05cm}   %调整图片标题与图距离
    \setlength{\belowcaptionskip}{-2cm}   %调整图片标题与下文距离
    \includegraphics[width=13cm]{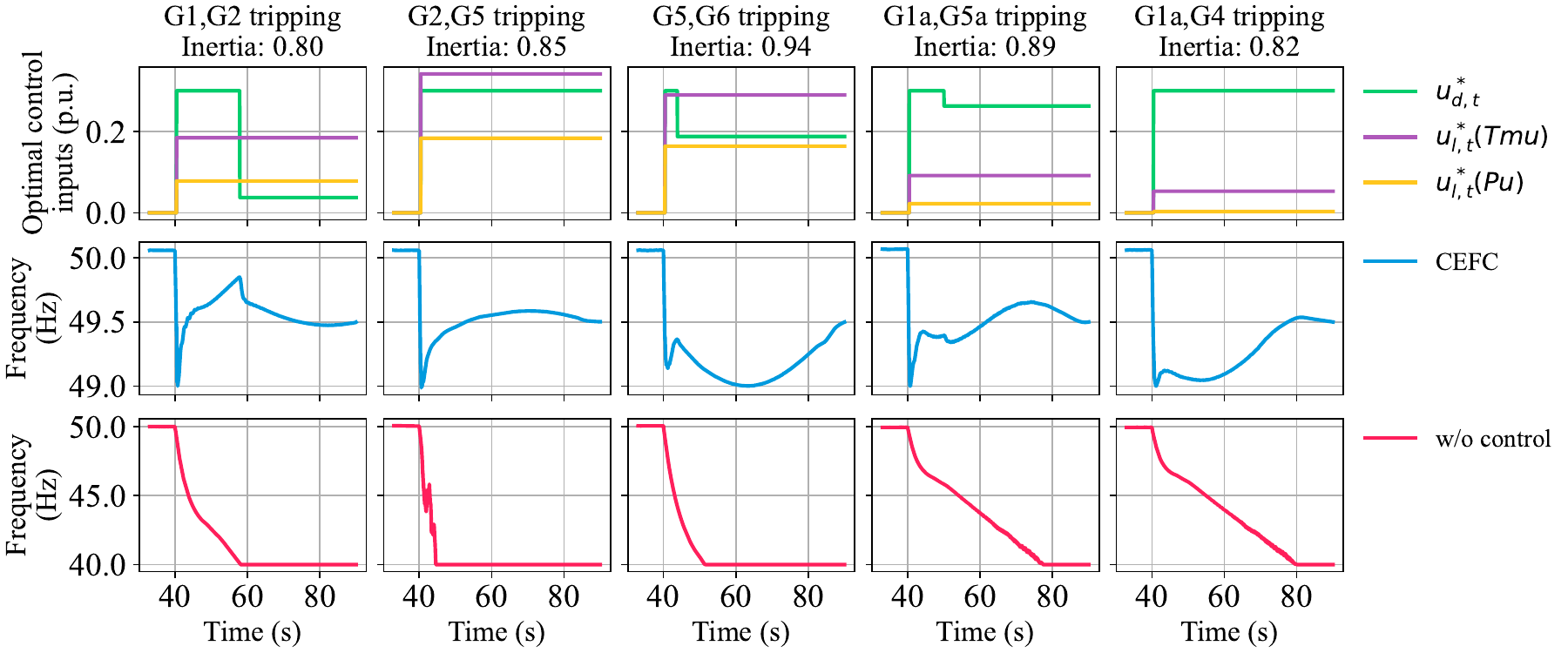}
    \caption{Control effectiveness of CEFC}\label{fig_optimal_control_KLS_large_fault_5_sub_figures}
\end{figure*}

In this subsection, we focus on investigating the impact of
prediction accuracy on control effectiveness. Fig. \ref{fig_optimal_control_KLS_large_fault_5_sub_figures} illustrates the system frequency response with and without CEFC under a relatively large power deficit for five subcases. In Fig. \ref{fig_optimal_control_KLS_large_fault_5_sub_figures}, $\boldsymbol{u}_{d,t}$ represents the total DC power modulation amount, $\boldsymbol{u}_{l,t} (T_{mu})$ represents the sum of the dynamic load shedding ratio (motor loads) of different nodes, and $\boldsymbol{u}_{l,t} (P_{u})$ represents the sum of the constant impedance load shedding amount of different nodes. The inertia of the equivalent unit in these five subcases is 0.8 p.u., 0.85 p.u., 0.94 p.u., 0.89 p.u., and 0.82 p.u. (with the original manufacturer data as the base value), respectively. The generator tripping faults are marked in the figure. In subcases (1)-(5), at the beginning of the fault, the power reference for the LCC-HVDCs reaches the upper limit of the DC active power for the sending end. The fast power adjustability of the LCC-HVDC systems is utilized to provide considerable power support and relieve the frequency moderation pressure of load shedding. In subcases (1), (3), and (4), after load shedding and EDCPS, the frequency gradually recovers. The adjustment of the DC power references decreases, reducing the control cost while maintaining the frequencies within the allowable deviation during the transient time and at the end of the control.

\begin{figure}[t] %可选参数 h t b p，代表允许图片出现的位置，h表示此处附近，t表示顶部，b表示底部，p表示单独一页，H表示固定此处
    \centering
    % \vspace{-0.2cm}  %调整图片与上文的垂直距离
    \setlength{\abovecaptionskip}{-0.05cm}   %调整图片标题与图距离
    \setlength{\belowcaptionskip}{-2cm}   %调整图片标题与下文距离
    \includegraphics[width=7cm]{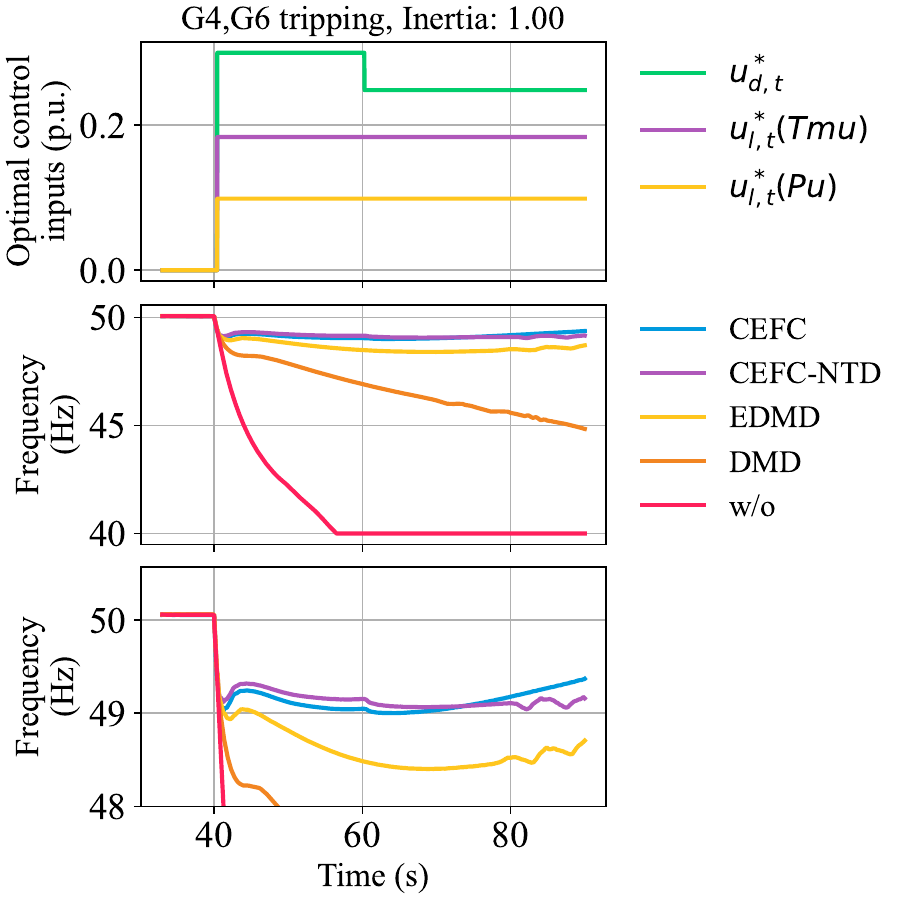}
    \caption{Comparison between CEFC and SOTAs on a scenario in the test set.}\label{optimal_control_large_fault_main_figure}
\end{figure}

Furthermore, we compared the control performance of the proposed method with SOTAs. Fig. \ref{optimal_control_large_fault_main_figure} displays a comparison of the control performance of CEFC and SOTAs when generators G4 and G6 are tripped. As seen from Fig. \ref{optimal_control_large_fault_main_figure}, due to its higher prediction accuracy, CEFC achieves frequency nadir and steady-state values closest to the set hard limits, thereby minimizing control costs while ensuring frequency safety. In contrast, DMD and EDMD overestimate the system frequency, resulting in insufficient control measures that cannot guarantee frequency safety. CEFC-NTD, with lower prediction accuracy than CEFC, yields a steady-state frequency slightly below the safety limit, while the frequency nadir is noticeably higher than the safety limit, failing to balance frequency safety and control cost reduction simultaneously.

\begin{figure}[t] %可选参数 h t b p，代表允许图片出现的位置，h表示此处附近，t表示顶部，b表示底部，p表示单独一页，H表示固定此处
    \centering
    % \vspace{-0.2cm}  %调整图片与上文的垂直距离
    \setlength{\abovecaptionskip}{-0.05cm}   %调整图片标题与图距离
    \setlength{\belowcaptionskip}{-2cm}   %调整图片标题与下文距离
    \includegraphics[width=9cm]{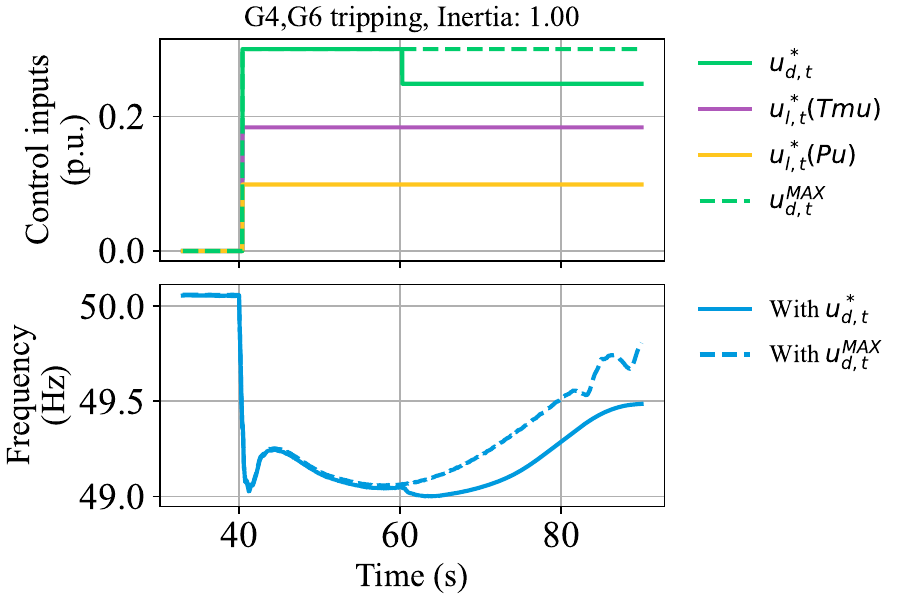}
    \caption{Comparison between the proposed closed-loop EDCPS and time-invariant EDCPS.}\label{optimal_control_lcc_compare}
\end{figure}

Finally, we demonstrate the necessity of the proposed closed-loop EDCPS in Eq. \eqref{Koopman_based_LQR_for_HVDC}. We aim to illustrate closed-loop EDCPS flexibly reduces the control cost while ensuring system frequency safety requirements. This is in comparison to maintaining the DC power reference at its adjustable limit. As shown in Fig. \ref{optimal_control_lcc_compare}, the blue dashed line represents the system trajectory when the DC power reference is maintained at its upper/lower limit. This control strategy can ensure system frequency safety, with the minimum frequency around 49 Hz and the final frequency at 49.8 Hz, clearly above the prescribed limit 49.5 Hz. The blue solid line depicts the system frequency trajectory under the proposed closed-loop EDCPS. After the system frequency passes the nadir, the control measure is reduced to ensure the minimization of the control cost while maintaining frequency safety.

\section{Conclusion}
This paper presents data-driven and non-preplanned EFC scheme, CEFC, which coordinates load shedding and EDCPS to ensure frequency safety. The approach combines open-loop load shedding with closed-loop EDCPS strategies, leveraging their respective strengths. Considering the one-shot action of load shedding, greater precision in control strategy is required. Sufficient conditions were established to ensure that system identification errors do not impact load shedding outcomes. Numerical case studies demonstrate that CEFC ensures both economic efficiency and safety in control strategies.
% \appendices
% \section{Effect of misrepresentation of Koopman eigenpairs}

% \section{}

% \section*{Appendix A}
% \vspace{-0.2cm}
% \section*{Proof of Proposition 1}

% \section*{Appendix B}
% \vspace{-0.2cm}
% \section*{Proof of Proposition 2}

\bibliographystyle{IEEEtran}
\bibliography{ref}

% Generated by IEEEtran.bst, version: 1.12 (2007/01/11)
\begin{thebibliography}{10}
\providecommand{\url}[1]{#1}
\csname url@samestyle\endcsname
\providecommand{\newblock}{\relax}
\providecommand{\bibinfo}[2]{#2}
\providecommand{\BIBentrySTDinterwordspacing}{\spaceskip=0pt\relax}
\providecommand{\BIBentryALTinterwordstretchfactor}{4}
\providecommand{\BIBentryALTinterwordspacing}{\spaceskip=\fontdimen2\font plus
\BIBentryALTinterwordstretchfactor\fontdimen3\font minus \fontdimen4\font\relax}
\providecommand{\BIBforeignlanguage}[2]{{%
\expandafter\ifx\csname l@#1\endcsname\relax
\typeout{** WARNING: IEEEtran.bst: No hyphenation pattern has been}%
\typeout{** loaded for the language `#1'. Using the pattern for}%
\typeout{** the default language instead.}%
\else
\language=\csname l@#1\endcsname
\fi
#2}}
\providecommand{\BIBdecl}{\relax}
\BIBdecl

\bibitem{9328133}
Y.~Liu, Y.~Song, Z.~Wang, and C.~Shen, ``Optimal emergency frequency control based on coordinated droop in multi-infeed hybrid ac-dc system,'' \emph{IEEE Transactions on Power Systems}, vol.~36, no.~4, pp. 3305--3316, 2021.

\bibitem{4495556}
R.~Li, S.~Bozhko, and G.~Asher, ``Frequency control design for offshore wind farm grid with lcc-hvdc link connection,'' \emph{IEEE Transactions on Power Electronics}, vol.~23, no.~3, pp. 1085--1092, 2008.

\bibitem{Bevrani2009}
H.~Bevrani, \emph{Robust Power System Frequency Control}.\hskip 1em plus 0.5em minus 0.4em\relax Berlin, Germany: Springer, 2009, vol.~85.

\bibitem{1137600}
G.~Karady and J.~Gu, ``A hybrid method for generator tripping,'' \emph{IEEE Transactions on Power Systems}, vol.~17, no.~4, pp. 1102--1107, 2002.

\bibitem{5546962}
U.~Rudez and R.~Mihalic, ``Monitoring the first frequency derivative to improve adaptive underfrequency load-shedding schemes,'' \emph{IEEE Transactions on Power Systems}, vol.~26, no.~2, pp. 839--846, 2011.

\bibitem{vittal2019power}
V.~Vittal, J.~D. McCalley, P.~M. Anderson, and A.~Fouad, \emph{Power system control and stability}.\hskip 1em plus 0.5em minus 0.4em\relax John Wiley \& Sons, 2019.

\bibitem{7581062}
Y.~Tofis, S.~Timotheou, and E.~Kyriakides, ``Minimal load shedding using the swing equation,'' \emph{IEEE Transactions on Power Systems}, vol.~32, no.~3, pp. 2466--2467, 2017.

\bibitem{10151932}
S.~Wu, F.~Liu, Z.~Wang, J.~Lin, and H.~Geng, ``Minimum energy demands of energy storages for fast frequency response: Formulation, solution, and implementation,'' \emph{IEEE Transactions on Power Systems}, vol.~39, no.~2, pp. 3615--3630, 2024.

\bibitem{9767658}
H.~Golpîra, H.~Bevrani, A.~Román~Messina, and B.~Francois, ``A data-driven under frequency load shedding scheme in power systems,'' \emph{IEEE Transactions on Power Systems}, vol.~38, no.~2, pp. 1138--1150, 2023.

\bibitem{8723612}
Q.~Wang, F.~Li, Y.~Tang, and Y.~Xu, ``Integrating model-driven and data-driven methods for power system frequency stability assessment and control,'' \emph{IEEE Transactions on Power Systems}, vol.~34, no.~6, pp. 4557--4568, 2019.

\bibitem{7337458}
U.~Rudez and R.~Mihalic, ``Wams-based underfrequency load shedding with short-term frequency prediction,'' \emph{IEEE Transactions on Power Delivery}, vol.~31, no.~4, pp. 1912--1920, 2016.

\bibitem{koopman1931hamiltonian}
B.~O. Koopman, ``Hamiltonian systems and transformation in hilbert space,'' \emph{Proceedings of the National Academy of Sciences}, vol.~17, no.~5, pp. 315--318, 1931.

\bibitem{KORDA2018297}
\BIBentryALTinterwordspacing
M.~Korda, Y.~Susuki, and I.~Mezić, ``Power grid transient stabilization using koopman model predictive control,'' \emph{IFAC-PapersOnLine}, vol.~51, no.~28, pp. 297--302, 2018, 10th IFAC Symposium on Control of Power and Energy Systems CPES 2018. [Online]. Available: \url{https://www.sciencedirect.com/science/article/pii/S2405896318334372}
\BIBentrySTDinterwordspacing

\bibitem{doi_10_1137_21M1401243}
\BIBentryALTinterwordspacing
S.~L. Brunton, M.~Budi\v{s}i\'{c}, E.~Kaiser, and J.~N. Kutz, ``Modern koopman theory for dynamical systems,'' \emph{SIAM Review}, vol.~64, no.~2, pp. 229--340, 2022. [Online]. Available: \url{https://doi.org/10.1137/21M1401243}
\BIBentrySTDinterwordspacing

\bibitem{Kaiser_2021}
\BIBentryALTinterwordspacing
E.~Kaiser, J.~N. Kutz, and S.~L. Brunton, ``Data-driven discovery of koopman eigenfunctions for control,'' \emph{Machine Learning: Science and Technology}, vol.~2, no.~3, p. 035023, jun 2021. [Online]. Available: \url{https://dx.doi.org/10.1088/2632-2153/abf0f5}
\BIBentrySTDinterwordspacing

\bibitem{cao2023data}
Q.~Cao, C.~Shen, and Y.~Liu, ``Data-driven emergency frequency control for multi-infeed hybrid ac-dc system,'' \emph{IEEE Transactions on Power Systems}, 2023.

\bibitem{xu2017design}
T.~Xu, G.~Li, Z.~Yu \emph{et~al.}, ``Design and application of emergency coordination control system for multi-infeed hvdc receiving-end system coping with frequency stability problem,'' \emph{Automation of Electric Power Systems}, vol.~41, no.~08, pp. 98--104, 2017.

\bibitem{10371235}
S.~Chandak, P.~Bhowmik, and P.~K. Rout, ``Load shedding strategy coordinated with storage device and d-statcom to enhance the microgrid stability,'' \emph{Protection and Control of Modern Power Systems}, vol.~4, no.~3, pp. 1--19, 2019.

\bibitem{santurino2022optimal}
P.~Santurino, L.~Sigrist, A.~Ortega, J.~Renedo, and E.~Lobato, ``Optimal coordinated design of under-frequency load shedding and energy storage systems,'' \emph{Electric Power Systems Research}, vol. 211, p. 108423, 2022.

\bibitem{10049715}
Y.~Liu, C.~Shen, Z.~Wang, and F.~Liu, ``Incentive mechanism design for emergency frequency control in multi-infeed hybrid ac-dc system,'' \emph{IEEE Transactions on Power Systems}, vol.~39, no.~1, pp. 1867--1880, 2024.

\bibitem{DL428-2010}
\emph{Technical Rules for Power System Automatic Under-Frequency Load Shedding}, China Power Industry Std. DL/T 428-2010, 2011.

\bibitem{CSEE-Benchmark2024}
``Electromechanical transient model - csee benchmark,'' \url{https://github.com/lbl-hub/CSEE-Benchmark}, accessed: 2024-01-15.

\bibitem{8582334}
Y.~Liu and Y.~Song, ``Modeling and simulation of hybrid ac-dc system on a cloud computing based simulation platform - cloudpss,'' in \emph{2018 2nd IEEE Conference on Energy Internet and Energy System Integration (EI2)}, 2018, pp. 1--6.

\end{thebibliography}
\end{document}